\documentclass[aps,prb,showpacs,twocolumn]{revtex4-1}

\usepackage{graphicx}
\usepackage{amsmath}
\usepackage{amsfonts}
\usepackage{amssymb}
\usepackage{epstopdf}

\usepackage[usenames,dvipsnames]{xcolor}
\newcommand{\bra}[1]{\langle #1|}
\newcommand{\ket}[1]{|#1\rangle}

\newcommand{\mean}[1]{\langle #1 \rangle}

\newcommand{\boldgreek}[1]{\ensuremath{\mbox{\boldmath$#1$}}}

\begin{document}

\author{Alexander Croy}
\email{alexander.croy@chalmers.se}

\affiliation
{Institute of Physics, Chemnitz University of Technology, D-09107 Chemnitz, Germany}

\affiliation{
Department of Applied Physics, Chalmers University of Technology, S-412 96 G\"oteborg, Sweden}

\author{Michael Schreiber}

\affiliation
{Institute of Physics, Chemnitz University of Technology, D-09107 Chemnitz, Germany}

\title
{Correlation-Strength Driven Anderson Metal-Insulator Transition}

\date{\today}

\begin{abstract}
The possibility of driving an Anderson metal-insulator transition in the presence of scale-free disorder by changing the correlation exponent is numerically investigated. 
We calculate the localization length for quasi-one-dimensional systems at fixed
energy and fixed disorder strength using a standard transfer matrix method. 
From a finite-size scaling analysis we extract the critical correlation exponent and the critical exponent characterizing the phase transition.
\end{abstract}


\maketitle
The Anderson model of localization \cite{And58} has been subject of intense study
over the past decades. In particular, the occurrence of a metal-insulator
phase transition (MIT) in three dimensions (3D) has attracted a lot of interest \cite{LeeR85,KraM93}. 
Theoretical studies of the MIT have focused mainly on situations with uncorrelated disorder \cite{BulSK87,KraM93,OhtSK99,RomS03}.
Therefore, one of the open questions in the field is the role of long-range correlated 
disorder in the Anderson MIT.

For uncorrelated disorder, the Anderson transition can be driven either by increasing
the disorder strength or by changing the Fermi energy \cite{KraM93}. In the former case,
for sufficiently strong disorder strength $W>W_{\rm c}(E)$ all electronic
states are exponentially localized, where the value $W_{\rm c}(E)$ depends on the
Fermi energy $E$. On the other hand, at fixed disorder strength states with $|E|<E_{\rm c}(W)$ are extended and otherwise localized.

The presence of correlations provides an additional possibility to achieve the MIT. Depending on the nature of the correlations, a transition may, in principle, be also driven by a change of the correlation strength or correlation length.
Such a scenario might be relevant in situations where the disorder is induced by a 
complex environment surrounding the system of interest.

In this Brief Report we study the possibility of a correlation-strength driven Anderson
MIT in 3D. We consider the case of scale-free disorder, which is characterized by a power-law with correlation exponent $\alpha$. We find at fixed energy {\it and} fixed disorder strength that the localization length behaves as
\begin{equation}\label{eq:Lambda}
  	\lambda(\alpha) \propto \left| \alpha_{\rm c} - \alpha \right|^{-\nu}\;,
\end{equation}
where the critical exponent $\nu$ depends on the values of $W$ and $E$. 
The obtained critical values $\alpha_{\rm c}$ are consistent with results 
for disorder and energy driven MITs 
in the presence of scale-free disorder \cite{CroCS11b}.

To study the influence of scale-free disorder on the Anderson MIT, we use
the usual tight-binding Hamiltonian in site representation \cite{And58,KraM93}
\begin{equation}
    \mathcal{H} = \sum_{\bf i} \varepsilon_{\bf i} \ket{{\bf i}}\bra{{\bf i}} 
    	- \sum_{\bf i\,j}\,t_{{\bf i j}}\,\ket{{\bf i}}\bra{{\bf j}}\;,
    \label{eq:AndersonHam}
\end{equation}
where $\ket{{\bf i}}$ denotes a localized state at lattice site ${\bf i}$. The
hopping matrix elements $t_{{\bf i j}}$ are restricted to nearest neighbors. As usual,
we set these elements to one and thereby fix the unit of energy.
The on-site potentials $\varepsilon_{\bf i}$ are taken as random numbers with a
Gaussian probability distribution. Specifically, we use random potentials with 
mean $\mean{\varepsilon_{\bf i}}=0$ and a correlation function of the form
	\begin{equation}\label{eq:PLCorrAsy}
		C(\boldgreek{\ell}) \equiv \mean{\varepsilon_{\bf i} \varepsilon_{{\bf i}+\boldgreek{\ell}}} \propto |\boldgreek{\ell}|^{-\alpha}\;,
	\end{equation}
where $\alpha$ is the correlation exponent. In the context of Anderson localization, this
correlation function has been used to study localization in the presence
of long-range correlations for one-dimensional \cite{IzrK99,deML98,RusHW98,Rus02,ShiNN04,Kay07,GarC09,CroCS11}, two-dimensional \cite{LiuCX99,LiuLL03,deMCLR04,MouLDM07,SanMLC07} and three-dimensional \cite{NdaRS04,CroCS11b} systems.

For the numerical calculations we generate the on-site potentials for systems 
of size $M\times M\times L$ using a modified Fourier filtering method ({FFM}) \cite{MakHSS96}. Additionally, we shift and scale the resulting random numbers to
have vanishing mean and variance $C(0)=W^2/12$. We focus on
quasi-one-dimensional systems with $L = 400000$ and $M=5,7,9,11$ and $13$.
The localization length $\lambda$ is calculated using a standard
transfer-matrix method (TMM) \cite{KraM93}. 
Monitoring the variance of the change of the Lyapunov exponent during the
TMM iterations gives a measure of the accuracy of the localization length
\cite{MacK81,*MacK83}.
We use a new seed
for each parameter combination ($E$, $W$, $\alpha$, $M$). Lastly, the critical
exponent and the critical correlation strength are obtained from a
finite-size scaling (FSS) analysis \cite{SleO99a}. We expand the one-parameter scaling law for the reduced localization length $\Lambda=\lambda/M$ into a Taylor series
\begin{equation}\label{eq:ScalingExpansion}
\Lambda(M,\tau) = \sum\limits^{n_{\rm I}}_{n=0} \phi^n M^{-n y} F_n(\chi M^{1/\nu})\;,
\end{equation}
where $\chi$ is a relevant scaling variable, $\phi$ is an irrelevant scaling variable, $y>0$ is the irrelevant scaling exponent and $\tau$ measures the distance from
the critical point. However, instead of using energy
or disorder strength to measure this distance, we
utilize the correlation exponent, i.e., $\tau=|\alpha-\alpha_{\rm c}|/\alpha_{\rm c}$. The functions $F_n$, $\chi$ and $\phi$ are 
further expanded up to order $n_{\rm R}$, $m_{\rm R}$ and $m_{\rm I}$, respectively.
Taking $n_{\rm I}>0$ allows us to consider corrections to scaling due 
to the finite size of the sample, which is reflected in a systematic shift of $\Lambda$ with $M$ in Eq.\ \eqref{eq:ScalingExpansion}.
Using a least squares fit of the expansion of the reduced localization length $\Lambda$ to the numerical data allows us to obtain the critical parameters.
Although one does {\em ad hoc} not expect the FSS analysis to be valid in the present case, we find that it is working surprisingly well,
as we will show in the following.

%
\begin{figure}[t]
   	\center
   	\includegraphics[width=.9\columnwidth]{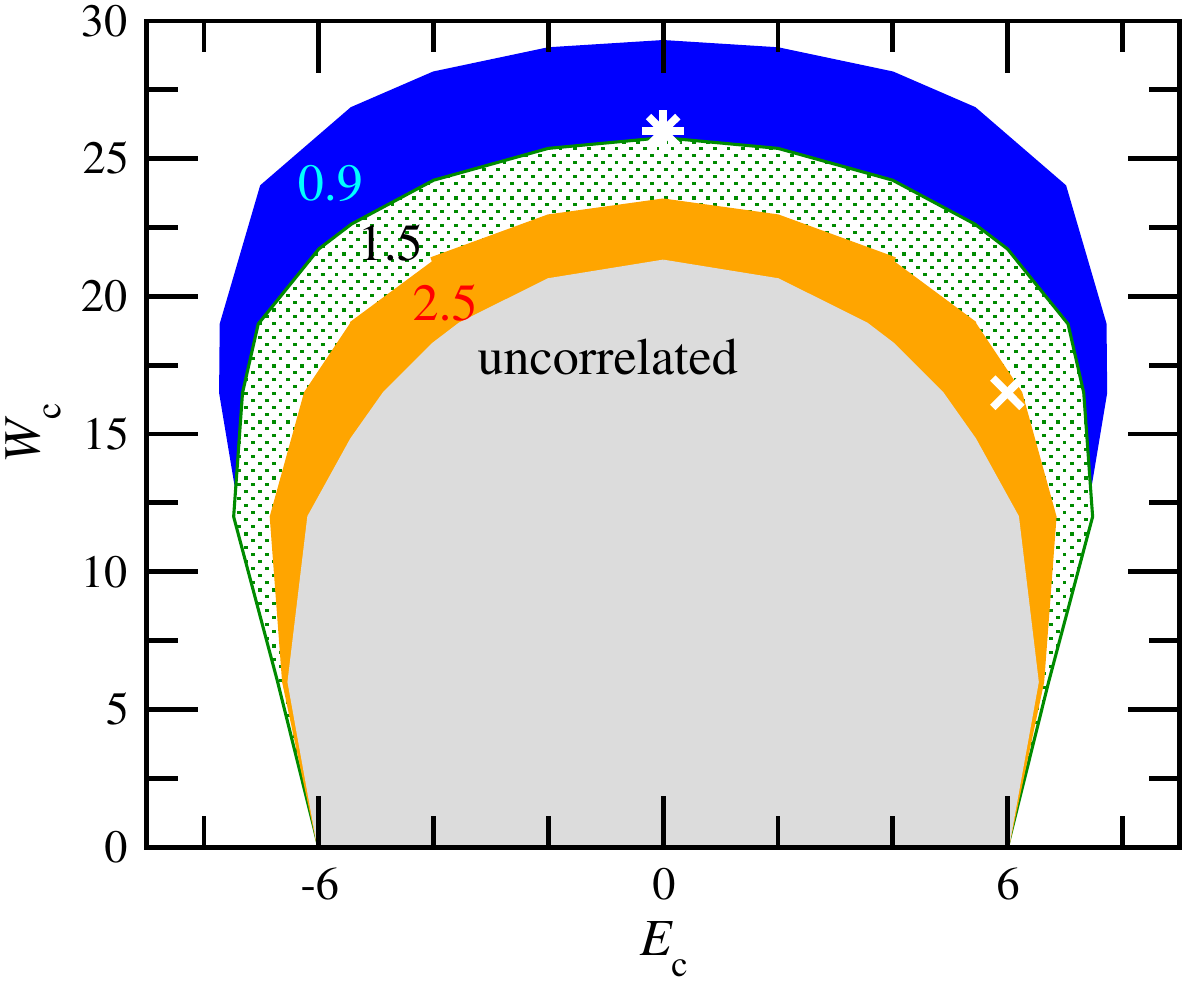}
   	\caption{Schematic phase diagram based on the results reported in 
		Ref.\ \cite{CroCS11b}. The phase-space points discussed in the text are 
		indicated by symbols ($*$ and $\times$).}
   	\label{fig:PhaseDia}
\end{figure}
We set $E={\rm const.}$ and $W={\rm const.}$ while varying $\alpha$. 
The chosen values of $E$ and $W$ are indicated in Fig.\ \ref{fig:PhaseDia}, which
shows a schematic phase diagram for the Anderson MIT in the presence of scale-free disorder. The position relative to the transition boundaries provides a first estimate
of the expected critical correlation exponent.

\begin{figure}[b!]
   	\center
   	\includegraphics[width=.9\columnwidth]{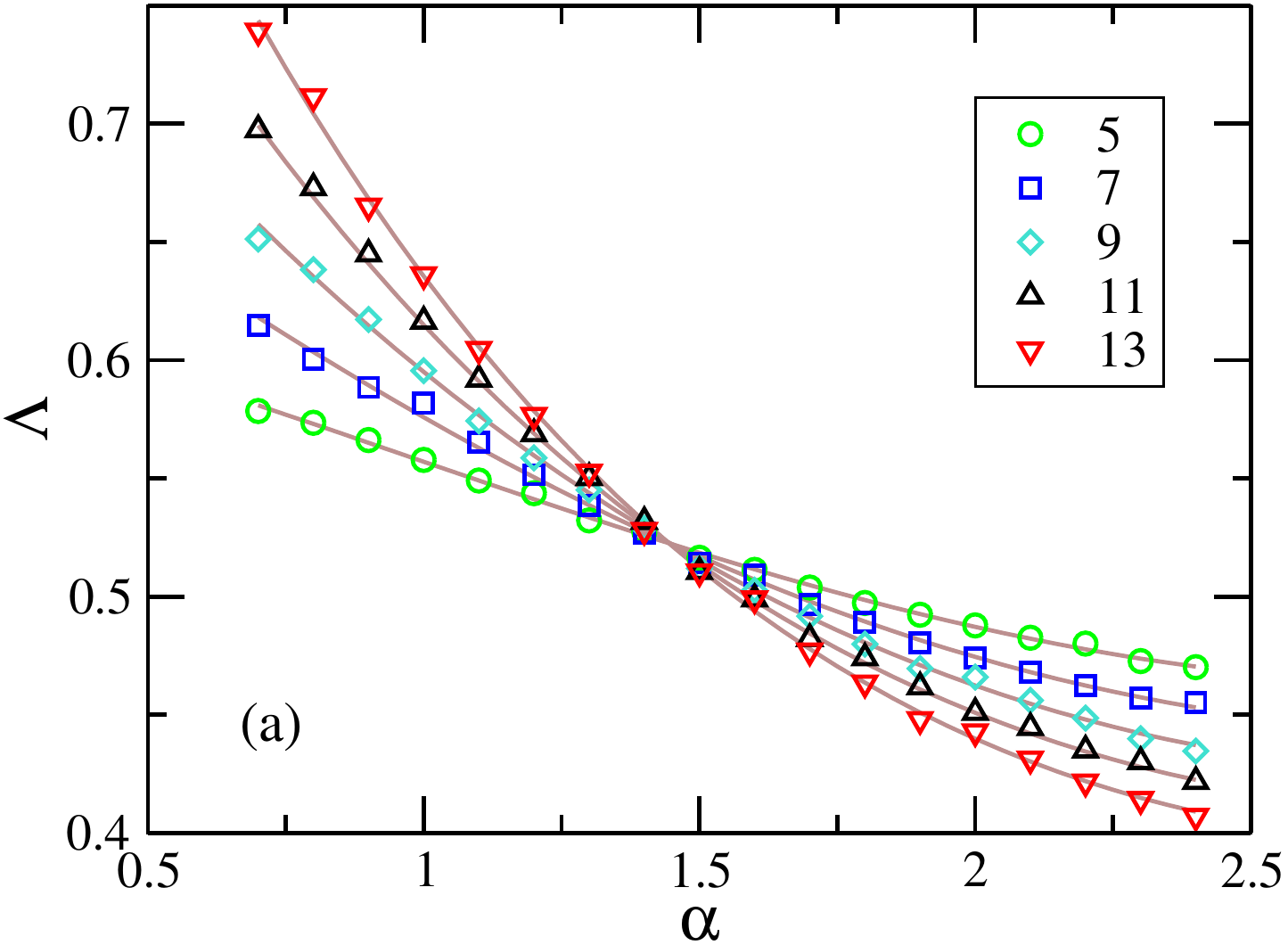}\vspace{2ex}
   	\includegraphics[width=.9\columnwidth]{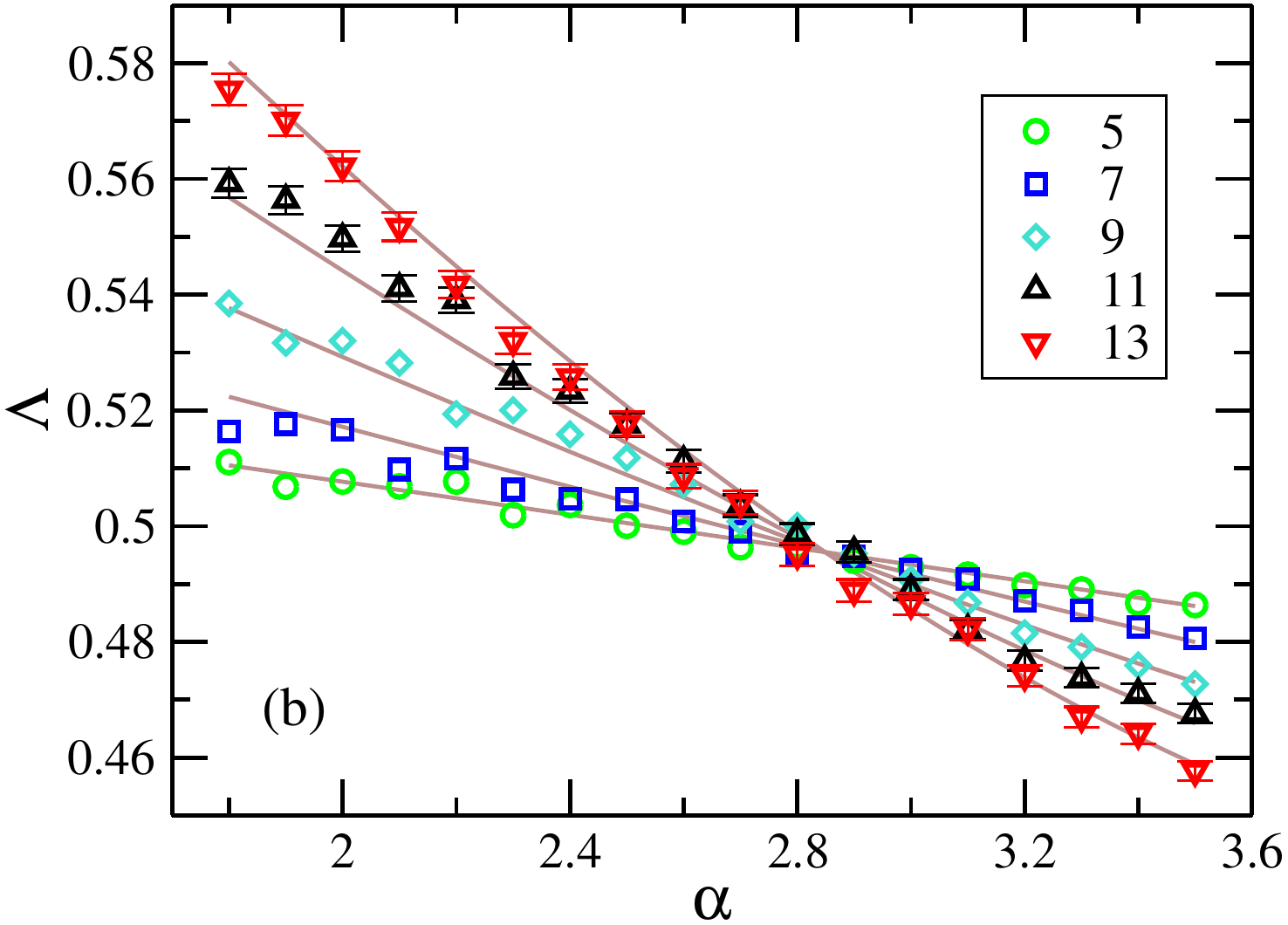}
   	\caption{Reduced localization length $\Lambda$ vs 	
   		correlation exponent $\alpha$.	Solid lines show 
		FSS fit to numerical data.
		(a) Taking corrections to scaling into account 
			 ($n_{\rm R}=2$, $n_{\rm I}=1$, $m_{\rm R}=2$, $m_{\rm I}=0$)
			 for $E=0.0, W=26.0$. 
		(b) Without taking corrections to scaling into account 
			 ($n_{\rm R}=2$, $n_{\rm I}=0$, $m_{\rm R}=2$, $m_{\rm I}=0$)
			 for $E=6.0, W=16.5$. 
	}
   	\label{fig:CorrTransE0-Corr}
	\label{fig:CorrTransE6-NoCorr}   	
\end{figure}

In Fig.\ \ref{fig:CorrTransE0-Corr} the reduced localization length
is shown in the vicinity of the band center, $E=0$, setting $W=26$. From the
dependence on the system size $M$, a clear transition can be seen. For small correlation
exponents ($\alpha< 1.5$) the reduced localization length increases with increasing
size $M$, while for large exponents ($\alpha > 1.5$) it decreases. The former is
characteristic for a metallic phase and the latter for an insulating phase.
The FSS procedure
yields for the critical correlation strength $\alpha_{\rm{c}}=1.44\pm0.04$, which
agrees very well with the value expected from the phase diagram. The critical exponent
is $\nu=0.98 \pm 0.09$ ($y=2.0\pm1.3$), which is different from the value $\nu_0=1.58\pm0.03$ obtained for uncorrelated disorder \cite{SleO99a} and from $\nu(\alpha=1.5) = 1.69 \pm 0.22$ reported for scale-free disorder \cite{CroCS11b}, both taken at $E=0$.

Also, for $E=6.0$ and $W=16.5$ we find a transition, as shown in
Fig.\ \ref{fig:CorrTransE6-NoCorr}. In this
case the critical value is found to be $\alpha_{\rm c}=2.85\pm0.03$, 
again consistent with the phase diagram in Fig.\ \ref{fig:PhaseDia}. The critical exponent is $\nu=0.62 \pm 0.03$, which is even smaller than the exponent found at the band center.

Qualitatively, the correlation-strength driven transition can be understood
by assuming an effective disorder strength, $W_{\rm eff}(\alpha)$, which depends on the correlation exponent.
An effective smoothening of the disorder potential has, for example, been observed for 1D systems, where the localization length in the band center increases for smaller correlation exponents \cite{IzrK99,CroCS11}. It is also in accordance with
the shift of the phase boundary towards higher energies and stronger disorder shown
in Fig.\ \ref{fig:PhaseDia}. Accordingly, the transition occurs when $W_{\rm eff}(\alpha_{\rm c})=W_{\rm c}(E)$. Close to the transition, the localization length would diverge 
according to $\lambda \propto |W_{\rm eff}(\alpha) - W_{\rm c}|^{-\nu_0} \propto |\alpha-\alpha_{\rm c}|^{-\nu_0}$, where we have expanded the effective disorder strength to first order, $W_{\rm eff}(\alpha)\approx W_{\rm c}
+ (\alpha-\alpha_{\rm c}) \partial W_{\rm eff}/\partial \alpha|_{\alpha_{\rm c}}$.
By construction this procedure yields the correct critical correlation strength, but
it does not explain the deviation of the observed critical exponents from the
universal value $\nu_0$. Provided the one-parameter scaling law holds in the
presence of long-range correlations, this discrepancy might also indicate that the FSS
method in the normally used form is not suitable to extract the critical exponent
in the present case.

In summary, we have studied the influence of scale-free disorder on the Anderson MIT
at fixed energy and fixed disorder strength. By varying the correlation exponent
we found an increasing reduced localization length for $\alpha<\alpha_{\rm c}$ and 
a decreasing reduced localization length for $\alpha>\alpha_{\rm c}$ when increasing
the system size. A FSS analysis yielded critical exponents which depend on the values
of $E$ and $W$ and are smaller than the universal value $\nu_0$ found
previously for uncorrelated disorder \cite{SleO99a}.


\begin{thebibliography}{25}%
\makeatletter
\providecommand \@ifxundefined [1]{%
 \@ifx{#1\undefined}
}%
\providecommand \@ifnum [1]{%
 \ifnum #1\expandafter \@firstoftwo
 \else \expandafter \@secondoftwo
 \fi
}%
\providecommand \@ifx [1]{%
 \ifx #1\expandafter \@firstoftwo
 \else \expandafter \@secondoftwo
 \fi
}%
\providecommand \natexlab [1]{#1}%
\providecommand \enquote  [1]{``#1''}%
\providecommand \bibnamefont  [1]{#1}%
\providecommand \bibfnamefont [1]{#1}%
\providecommand \citenamefont [1]{#1}%
\providecommand \href@noop [0]{\@secondoftwo}%
\providecommand \href [0]{\begingroup \@sanitize@url \@href}%
\providecommand \@href[1]{\@@startlink{#1}\@@href}%
\providecommand \@@href[1]{\endgroup#1\@@endlink}%
\providecommand \@sanitize@url [0]{\catcode `\\12\catcode `\$12\catcode
  `\&12\catcode `\#12\catcode `\^12\catcode `\_12\catcode `\%12\relax}%
\providecommand \@@startlink[1]{}%
\providecommand \@@endlink[0]{}%
\providecommand \url  [0]{\begingroup\@sanitize@url \@url }%
\providecommand \@url [1]{\endgroup\@href {#1}{\urlprefix }}%
\providecommand \urlprefix  [0]{URL }%
\providecommand \Eprint [0]{\href }%
\providecommand \doibase [0]{http://dx.doi.org/}%
\providecommand \selectlanguage [0]{\@gobble}%
\providecommand \bibinfo  [0]{\@secondoftwo}%
\providecommand \bibfield  [0]{\@secondoftwo}%
\providecommand \translation [1]{[#1]}%
\providecommand \BibitemOpen [0]{}%
\providecommand \bibitemStop [0]{}%
\providecommand \bibitemNoStop [0]{.\EOS\space}%
\providecommand \EOS [0]{\spacefactor3000\relax}%
\providecommand \BibitemShut  [1]{\csname bibitem#1\endcsname}%
\let\auto@bib@innerbib\@empty
\bibitem [{\citenamefont {Anderson}(1958)}]{And58}%
  \BibitemOpen
  \bibfield  {author} {\bibinfo {author} {\bibfnamefont {P.~W.}\ \bibnamefont
  {Anderson}},\ }\href@noop {} {\bibfield  {journal} {\bibinfo  {journal}
  {Phys. Rev.}\ }\textbf {\bibinfo {volume} {109}},\ \bibinfo {pages} {1492}
  (\bibinfo {year} {1958})}\BibitemShut {NoStop}%
\bibitem [{\citenamefont {Lee}\ and\ \citenamefont
  {Ramakrishnan}(1985)}]{LeeR85}%
  \BibitemOpen
  \bibfield  {author} {\bibinfo {author} {\bibfnamefont {P.~A.}\ \bibnamefont
  {Lee}}\ and\ \bibinfo {author} {\bibfnamefont {T.~V.}\ \bibnamefont
  {Ramakrishnan}},\ }\href@noop {} {\bibfield  {journal} {\bibinfo  {journal}
  {Rev. Mod. Phys.}\ }\textbf {\bibinfo {volume} {57}},\ \bibinfo {pages} {287}
  (\bibinfo {year} {1985})}\BibitemShut {NoStop}%
\bibitem [{\citenamefont {Kramer}\ and\ \citenamefont
  {MacKinnon}(1993)}]{KraM93}%
  \BibitemOpen
  \bibfield  {author} {\bibinfo {author} {\bibfnamefont {B.}~\bibnamefont
  {Kramer}}\ and\ \bibinfo {author} {\bibfnamefont {A.}~\bibnamefont
  {MacKinnon}},\ }\href {\doibase 10.1088/0034-4885/56/12/001} {\bibfield
  {journal} {\bibinfo  {journal} {Rep. Prog. Phys.}\ }\textbf {\bibinfo
  {volume} {56}},\ \bibinfo {pages} {1469} (\bibinfo {year}
  {1993})}\BibitemShut {NoStop}%
\bibitem [{\citenamefont {Bulka}\ \emph {et~al.}(1987)\citenamefont {Bulka},
  \citenamefont {Schreiber},\ and\ \citenamefont {Kramer}}]{BulSK87}%
  \BibitemOpen
  \bibfield  {author} {\bibinfo {author} {\bibfnamefont {B.}~\bibnamefont
  {Bulka}}, \bibinfo {author} {\bibfnamefont {M.}~\bibnamefont {Schreiber}}, \
  and\ \bibinfo {author} {\bibfnamefont {B.}~\bibnamefont {Kramer}},\
  }\href@noop {} {\bibfield  {journal} {\bibinfo  {journal} {Z. Phys. B}\
  }\textbf {\bibinfo {volume} {66}},\ \bibinfo {pages} {21} (\bibinfo {year}
  {1987})}\BibitemShut {NoStop}%
\bibitem [{\citenamefont {Ohtsuki}\ \emph {et~al.}(1999)\citenamefont
  {Ohtsuki}, \citenamefont {Slevin},\ and\ \citenamefont
  {Kawarabayashi}}]{OhtSK99}%
  \BibitemOpen
  \bibfield  {author} {\bibinfo {author} {\bibfnamefont {T.}~\bibnamefont
  {Ohtsuki}}, \bibinfo {author} {\bibfnamefont {K.}~\bibnamefont {Slevin}}, \
  and\ \bibinfo {author} {\bibfnamefont {T.}~\bibnamefont {Kawarabayashi}},\
  }\href@noop {} {\bibfield  {journal} {\bibinfo  {journal} {Ann. Phys.
  (Leipzig)}\ }\textbf {\bibinfo {volume} {8}},\ \bibinfo {pages} {655}
  (\bibinfo {year} {1999})}\BibitemShut {NoStop}%
\bibitem [{\citenamefont {{R\"{o}mer}}\ and\ \citenamefont
  {Schreiber}(2003)}]{RomS03}%
  \BibitemOpen
  \bibfield  {author} {\bibinfo {author} {\bibfnamefont {R.~A.}\ \bibnamefont
  {{R\"{o}mer}}}\ and\ \bibinfo {author} {\bibfnamefont {M.}~\bibnamefont
  {Schreiber}},\ }\enquote {\bibinfo {title} {The {A}nderson transition and its
  ramifications --- localisation, quantum interference, and interactions},}\ \
  (\bibinfo  {publisher} {Springer},\ \bibinfo {address} {Berlin},\ \bibinfo
  {year} {2003})\ Chap.\ \bibinfo {chapter} {Numerical investigations of
  scaling at the Anderson transition}, pp.\ \bibinfo {pages}
  {3--19}\BibitemShut {NoStop}%
\bibitem [{\citenamefont {Croy}\ \emph
  {et~al.}(2011{\natexlab{a}})\citenamefont {Croy}, \citenamefont {Cain},\ and\
  \citenamefont {Schreiber}}]{CroCS11b}%
  \BibitemOpen
  \bibfield  {author} {\bibinfo {author} {\bibfnamefont {A.}~\bibnamefont
  {Croy}}, \bibinfo {author} {\bibfnamefont {P.}~\bibnamefont {Cain}}, \ and\
  \bibinfo {author} {\bibfnamefont {M.}~\bibnamefont {Schreiber}},\ }\href@noop
  {} {\enquote {\bibinfo {title} {The role of power-law correlated disorder in
  the {A}nderson metal-insulator transition},}\ } (\bibinfo {year}
  {2011}{\natexlab{a}}),\ \bibinfo {note} {arXiv:1112.4469v1}\BibitemShut
  {NoStop}%
\bibitem [{\citenamefont {Izrailev}\ and\ \citenamefont
  {Krokhin}(1999)}]{IzrK99}%
  \BibitemOpen
  \bibfield  {author} {\bibinfo {author} {\bibfnamefont {F.~M.}\ \bibnamefont
  {Izrailev}}\ and\ \bibinfo {author} {\bibfnamefont {A.~A.}\ \bibnamefont
  {Krokhin}},\ }\href@noop {} {\bibfield  {journal} {\bibinfo  {journal} {Phys.
  Rev. Lett.}\ }\textbf {\bibinfo {volume} {82}},\ \bibinfo {pages} {4062}
  (\bibinfo {year} {1999})}\BibitemShut {NoStop}%
\bibitem [{\citenamefont {de~Moura}\ and\ \citenamefont {Lyra}(1998)}]{deML98}%
  \BibitemOpen
  \bibfield  {author} {\bibinfo {author} {\bibfnamefont {F.~A. B.~F.}\
  \bibnamefont {de~Moura}}\ and\ \bibinfo {author} {\bibfnamefont {M.~L.}\
  \bibnamefont {Lyra}},\ }\href {\doibase 10.1103/PhysRevLett.81.3735}
  {\bibfield  {journal} {\bibinfo  {journal} {Phys. Rev. Lett.}\ }\textbf
  {\bibinfo {volume} {81}},\ \bibinfo {pages} {3735} (\bibinfo {year}
  {1998})}\BibitemShut {NoStop}%
\bibitem [{\citenamefont {Russ}\ \emph {et~al.}(1998)\citenamefont {Russ},
  \citenamefont {Havlin},\ and\ \citenamefont {Webman}}]{RusHW98}%
  \BibitemOpen
  \bibfield  {author} {\bibinfo {author} {\bibfnamefont {S.}~\bibnamefont
  {Russ}}, \bibinfo {author} {\bibfnamefont {S.}~\bibnamefont {Havlin}}, \ and\
  \bibinfo {author} {\bibfnamefont {I.}~\bibnamefont {Webman}},\ }\href@noop {}
  {\bibfield  {journal} {\bibinfo  {journal} {Phil. Mag. B}\ }\textbf {\bibinfo
  {volume} {77}},\ \bibinfo {pages} {1449} (\bibinfo {year}
  {1998})}\BibitemShut {NoStop}%
\bibitem [{\citenamefont {Russ}(2002)}]{Rus02}%
  \BibitemOpen
  \bibfield  {author} {\bibinfo {author} {\bibfnamefont {S.}~\bibnamefont
  {Russ}},\ }\href {\doibase 10.1103/PhysRevB.66.012204} {\bibfield  {journal}
  {\bibinfo  {journal} {Phys. Rev. B}\ }\textbf {\bibinfo {volume} {66}},\
  \bibinfo {pages} {012204} (\bibinfo {year} {2002})}\BibitemShut {NoStop}%
\bibitem [{\citenamefont {Shima}\ \emph {et~al.}(2004)\citenamefont {Shima},
  \citenamefont {Nomura},\ and\ \citenamefont {Nakayama}}]{ShiNN04}%
  \BibitemOpen
  \bibfield  {author} {\bibinfo {author} {\bibfnamefont {H.}~\bibnamefont
  {Shima}}, \bibinfo {author} {\bibfnamefont {T.}~\bibnamefont {Nomura}}, \
  and\ \bibinfo {author} {\bibfnamefont {T.}~\bibnamefont {Nakayama}},\ }\href
  {\doibase 10.1103/PhysRevB.70.075116} {\bibfield  {journal} {\bibinfo
  {journal} {Phys. Rev. B}\ }\textbf {\bibinfo {volume} {70}},\ \bibinfo
  {pages} {075116} (\bibinfo {year} {2004})}\BibitemShut {NoStop}%
\bibitem [{\citenamefont {Kaya}(2007)}]{Kay07}%
  \BibitemOpen
  \bibfield  {author} {\bibinfo {author} {\bibfnamefont {T.}~\bibnamefont
  {Kaya}},\ }\href {http://dx.doi.org/10.1140/epjb/e2007-00036-4} {\bibfield
  {journal} {\bibinfo  {journal} {Eur. Phys. J. B}\ }\textbf {\bibinfo {volume}
  {55}},\ \bibinfo {pages} {49} (\bibinfo {year} {2007})}\BibitemShut {NoStop}%
\bibitem [{\citenamefont {Garc\'{\i}a-Garc\'{\i}a}\ and\ \citenamefont
  {Cuevas}(2009)}]{GarC09}%
  \BibitemOpen
  \bibfield  {author} {\bibinfo {author} {\bibfnamefont {A.~M.}\ \bibnamefont
  {Garc\'{\i}a-Garc\'{\i}a}}\ and\ \bibinfo {author} {\bibfnamefont
  {E.}~\bibnamefont {Cuevas}},\ }\href {\doibase 10.1103/PhysRevB.79.073104}
  {\bibfield  {journal} {\bibinfo  {journal} {Phys. Rev. B}\ }\textbf {\bibinfo
  {volume} {79}},\ \bibinfo {pages} {073104} (\bibinfo {year}
  {2009})}\BibitemShut {NoStop}%
\bibitem [{\citenamefont {Croy}\ \emph
  {et~al.}(2011{\natexlab{b}})\citenamefont {Croy}, \citenamefont {Cain},\ and\
  \citenamefont {Schreiber}}]{CroCS11}%
  \BibitemOpen
  \bibfield  {author} {\bibinfo {author} {\bibfnamefont {A.}~\bibnamefont
  {Croy}}, \bibinfo {author} {\bibfnamefont {P.}~\bibnamefont {Cain}}, \ and\
  \bibinfo {author} {\bibfnamefont {M.}~\bibnamefont {Schreiber}},\ }\href@noop
  {} {\bibfield  {journal} {\bibinfo  {journal} {Eur. Phys. J. B}\ }\textbf
  {\bibinfo {volume} {82}},\ \bibinfo {pages} {107} (\bibinfo {year}
  {2011}{\natexlab{b}})}\BibitemShut {NoStop}%
\bibitem [{\citenamefont {Liu}\ \emph {et~al.}(1999)\citenamefont {Liu},
  \citenamefont {Chen},\ and\ \citenamefont {Xiong}}]{LiuCX99}%
  \BibitemOpen
  \bibfield  {author} {\bibinfo {author} {\bibfnamefont {W.}~\bibnamefont
  {Liu}}, \bibinfo {author} {\bibfnamefont {T.}~\bibnamefont {Chen}}, \ and\
  \bibinfo {author} {\bibfnamefont {S.}~\bibnamefont {Xiong}},\ }\href@noop {}
  {\bibfield  {journal} {\bibinfo  {journal} {J. Phys. C}\ }\textbf {\bibinfo
  {volume} {11}},\ \bibinfo {pages} {6883} (\bibinfo {year}
  {1999})}\BibitemShut {NoStop}%
\bibitem [{\citenamefont {Liu}\ \emph {et~al.}(2003)\citenamefont {Liu},
  \citenamefont {Liu},\ and\ \citenamefont {Lei}}]{LiuLL03}%
  \BibitemOpen
  \bibfield  {author} {\bibinfo {author} {\bibfnamefont {W.-S.}\ \bibnamefont
  {Liu}}, \bibinfo {author} {\bibfnamefont {S.}~\bibnamefont {Liu}}, \ and\
  \bibinfo {author} {\bibfnamefont {X.}~\bibnamefont {Lei}},\ }\href {\doibase
  10.1140/epjb/e2003-00169-4} {\bibfield  {journal} {\bibinfo  {journal} {Eur.
  Phys. J. B}\ }\textbf {\bibinfo {volume} {33}},\ \bibinfo {pages} {293}
  (\bibinfo {year} {2003})}\BibitemShut {NoStop}%
\bibitem [{\citenamefont {de~Moura}\ \emph {et~al.}(2004)\citenamefont
  {de~Moura}, \citenamefont {Coutinho-Filho}, \citenamefont {Lyra},\ and\
  \citenamefont {Raposo}}]{deMCLR04}%
  \BibitemOpen
  \bibfield  {author} {\bibinfo {author} {\bibfnamefont {F.~A. B.~F.}\
  \bibnamefont {de~Moura}}, \bibinfo {author} {\bibfnamefont {M.~D.}\
  \bibnamefont {Coutinho-Filho}}, \bibinfo {author} {\bibfnamefont {M.~L.}\
  \bibnamefont {Lyra}}, \ and\ \bibinfo {author} {\bibfnamefont {E.~P.}\
  \bibnamefont {Raposo}},\ }\href
  {http://stacks.iop.org/0295-5075/66/i=4/a=585} {\bibfield  {journal}
  {\bibinfo  {journal} {Europhys. Lett.}\ }\textbf {\bibinfo {volume} {66}},\
  \bibinfo {pages} {585} (\bibinfo {year} {2004})}\BibitemShut {NoStop}%
\bibitem [{\citenamefont {de~Moura}\ \emph {et~al.}(2007)\citenamefont
  {de~Moura}, \citenamefont {Lyra}, \citenamefont {Dom\'{i}nguez-Adame},\ and\
  \citenamefont {Malyshev}}]{MouLDM07}%
  \BibitemOpen
  \bibfield  {author} {\bibinfo {author} {\bibfnamefont {F.~A. B.~F.}\
  \bibnamefont {de~Moura}}, \bibinfo {author} {\bibfnamefont {M.~L.}\
  \bibnamefont {Lyra}}, \bibinfo {author} {\bibfnamefont {F.}~\bibnamefont
  {Dom\'{i}nguez-Adame}}, \ and\ \bibinfo {author} {\bibfnamefont {V.~A.}\
  \bibnamefont {Malyshev}},\ }\href
  {http://stacks.iop.org/0953-8984/19/i=5/a=056204} {\bibfield  {journal}
  {\bibinfo  {journal} {J. Phys. C}\ }\textbf {\bibinfo {volume} {19}},\
  \bibinfo {pages} {056204} (\bibinfo {year} {2007})}\BibitemShut {NoStop}%
\bibitem [{\citenamefont {dos Santos}\ \emph {et~al.}(2007)\citenamefont {dos
  Santos}, \citenamefont {de~Moura}, \citenamefont {Lyra},\ and\ \citenamefont
  {Coutinho-Filho}}]{SanMLC07}%
  \BibitemOpen
  \bibfield  {author} {\bibinfo {author} {\bibfnamefont {I.~F.}\ \bibnamefont
  {dos Santos}}, \bibinfo {author} {\bibfnamefont {F.~A. B.~F.}\ \bibnamefont
  {de~Moura}}, \bibinfo {author} {\bibfnamefont {M.~L.}\ \bibnamefont {Lyra}},
  \ and\ \bibinfo {author} {\bibfnamefont {M.~D.}\ \bibnamefont
  {Coutinho-Filho}},\ }\href {http://stacks.iop.org/0953-8984/19/i=47/a=476213}
  {\bibfield  {journal} {\bibinfo  {journal} {J. Phys. C}\ }\textbf {\bibinfo
  {volume} {19}},\ \bibinfo {pages} {476213} (\bibinfo {year}
  {2007})}\BibitemShut {NoStop}%
\bibitem [{\citenamefont {Ndawana}\ \emph {et~al.}(2004)\citenamefont
  {Ndawana}, \citenamefont {{R\"{o}mer}},\ and\ \citenamefont
  {Schreiber}}]{NdaRS04}%
  \BibitemOpen
  \bibfield  {author} {\bibinfo {author} {\bibfnamefont {M.~L.}\ \bibnamefont
  {Ndawana}}, \bibinfo {author} {\bibfnamefont {R.~A.}\ \bibnamefont
  {{R\"{o}mer}}}, \ and\ \bibinfo {author} {\bibfnamefont {M.}~\bibnamefont
  {Schreiber}},\ }\href@noop {} {\bibfield  {journal} {\bibinfo  {journal}
  {Europhys. Lett.}\ }\textbf {\bibinfo {volume} {68}},\ \bibinfo {pages} {678}
  (\bibinfo {year} {2004})}\BibitemShut {NoStop}%
\bibitem [{\citenamefont {Makse}\ \emph {et~al.}(1996)\citenamefont {Makse},
  \citenamefont {Havlin}, \citenamefont {Schwartz},\ and\ \citenamefont
  {Stanley}}]{MakHSS96}%
  \BibitemOpen
  \bibfield  {author} {\bibinfo {author} {\bibfnamefont {H.~A.}\ \bibnamefont
  {Makse}}, \bibinfo {author} {\bibfnamefont {S.}~\bibnamefont {Havlin}},
  \bibinfo {author} {\bibfnamefont {M.}~\bibnamefont {Schwartz}}, \ and\
  \bibinfo {author} {\bibfnamefont {H.~E.}\ \bibnamefont {Stanley}},\
  }\href@noop {} {\bibfield  {journal} {\bibinfo  {journal} {Phys. Rev. E}\
  }\textbf {\bibinfo {volume} {53}},\ \bibinfo {pages} {5445} (\bibinfo {year}
  {1996})}\BibitemShut {NoStop}%
\bibitem [{\citenamefont {MacKinnon}\ and\ \citenamefont
  {Kramer}(1981)}]{MacK81}%
  \BibitemOpen
  \bibfield  {author} {\bibinfo {author} {\bibfnamefont {A.}~\bibnamefont
  {MacKinnon}}\ and\ \bibinfo {author} {\bibfnamefont {B.}~\bibnamefont
  {Kramer}},\ }\href {\doibase 10.1103/PhysRevLett.47.1546} {\bibfield
  {journal} {\bibinfo  {journal} {Phys. Rev. Lett.}\ }\textbf {\bibinfo
  {volume} {47}},\ \bibinfo {pages} {1546} (\bibinfo {year}
  {1981})}\BibitemShut {NoStop}%
\bibitem [{\citenamefont {MacKinnon}\ and\ \citenamefont
  {Kramer}(1983)}]{MacK83}%
  \BibitemOpen
  \bibfield  {author} {\bibinfo {author} {\bibfnamefont {A.}~\bibnamefont
  {MacKinnon}}\ and\ \bibinfo {author} {\bibfnamefont {B.}~\bibnamefont
  {Kramer}},\ }\href {\doibase 10.1007/BF01578242} {\bibfield  {journal}
  {\bibinfo  {journal} {Z. Phys. B}\ }\textbf {\bibinfo {volume} {53}},\
  \bibinfo {pages} {1} (\bibinfo {year} {1983})}\BibitemShut {NoStop}%
\bibitem [{\citenamefont {Slevin}\ and\ \citenamefont
  {Ohtsuki}(1999)}]{SleO99a}%
  \BibitemOpen
  \bibfield  {author} {\bibinfo {author} {\bibfnamefont {K.}~\bibnamefont
  {Slevin}}\ and\ \bibinfo {author} {\bibfnamefont {T.}~\bibnamefont
  {Ohtsuki}},\ }\href@noop {} {\bibfield  {journal} {\bibinfo  {journal} {Phys.
  Rev. Lett.}\ }\textbf {\bibinfo {volume} {82}},\ \bibinfo {pages} {382}
  (\bibinfo {year} {1999})}\BibitemShut {NoStop}%
\end{thebibliography}
%

\end{document}